\documentclass[prb,aps,twocolumn,showkeys] {revtex4}

\usepackage{xspace}

\usepackage[dvips]{graphicx}
\usepackage{epsfig}
\usepackage{amssymb}

\usepackage{soul,color}

\begin{document}

\title{Beyond Moore's technologies: operation principles of a superconductor alternative}

\author{I. I. Soloviev$^{1,2}$}
\email[]{isol@phys.msu.ru}
\author{N. V. Klenov$^{1,2,3}$}
\author{S. V. Bakurskiy$^{1}$}
\author{M. Yu. Kupriyanov$^{1,4}$}
\author{A. L. Gudkov$^{5}$}
\author{A. S. Sidorenko$^{4,6}$}

\affiliation{$^{1}$Lomonosov Moscow State University, Skobeltsyn
Institute of Nuclear Physics, 119991, Moscow, Russia}
\affiliation{$^{2}$Moscow Technological University (MIREA), 119454,
Moscow, Russia} \affiliation{$^{3}$All-Russian Research Institute of
Automatics n.a. N.L. Dukhov (VNIIA), 127055, Moscow, Russia}
\affiliation{$^{4}$Solid State Physics Department, Kazan Federal
University, 420008, Kazan, Russia} \affiliation{$^{5}$Lukin
Scientific Research Institute of Physical Problems, 124460,
Zelenograd, Moscow, Russia} \affiliation{$^{6}$Ghitu Institute of
Electronic Engineering and Nanotechnologies ASM, Chisinau, Moldova}

\date{\today}



\begin{abstract}
The predictions of Moore's law are considered by experts to be valid
until 2020 giving rise to ``post-Moore's'' technologies afterwards.
Energy efficiency is one of the major challenges in high-performance
computing that should be answered. Superconductor digital technology
is a promising post-Moore's alternative for the development of
supercomputers. In this paper, we consider operation principles of
an energy-efficient superconductor logic and memory circuits with a
short retrospective review of their evolution. We analyze their
shortcomings in respect to computer circuits design. Possible ways
of further research are outlined.
\end{abstract}

\maketitle

\section*{Introduction}
World's largest chipmaker Intel ``has signaled a slowing of Moore's
Law'' \cite{MLD1}. The company has decided to increase the time
between future generations of chips.  ``A technology roadmap for
Moore's Law maintained by an industry group, including the world's
largest chip makers, is being scrapped'' \cite{MLD}.

Three years ago Bob Colwell (former Intel chief IA-32 architect on
the Pentium Pro, Pentium II, Pentium III, and Pentium IV) described
stagnation of semiconductor technology in the following sentences
\cite{10}:

-- Officially Moore's Law ends in 2020 at 7~nm, but nobody cares,
because 11~nm isn't any better than 14~nm, which was only marginally
better than 22~nm.

-- With Dennard scaling already dead since 2004, and thermal
dissipation issues thoroughly constrain the integration density --
effectively ending the multicore era: ``Dark Silicon'' problem (only
part of available cores can be run simultaneously).

The mentioned fundamental changes are most clearly manifested in
supercomputer industry. Energy efficiency becomes a crucial
parameter constraining its headway \cite{1,2,3}. Power consumption
level of the most powerful modern supercomputer Sunway TaihuLight
\cite{4} is as high as 15.4~MW. It corresponds to peak performance
of 93 petaflops (1 petaflops is $10^{15}$ floating point operations
per second). Power consumption level of the next generation --
exaflops ($10^{18}$ flops) supercomputers is predicted \cite{5} to
be in the sub-GW level. It is comparable to power generated by a
small powerplant and results in an unreasonable bill of hundreds of
million dollars per year.

Following roadmap \cite{6}, goal power consumption level of exaflops
supercomputer should be of the order of $\sim 20$~MW. It corresponds
to energy efficiency of 20~pJ/flop or 50~Gflops/W. Unfortunately,
energy efficiency of modern supercomputers is an order less than
required. For example, the energy efficiency \cite{4} of the Sunway
TaihuLight is 6~Gflops/W. It is understood that besides other issues
of exaflops computer like large space and complex cooling
infrastructure, the energy efficiency makes the next step in high
performance computing to be extraordinarily difficult, even with
planned advances in complementary metal-oxide-semiconductor (CMOS)
technology \cite{3DP}.

It is worth to mention that low energy efficiency leads to high
power consumption and also limits clock frequency at the level of 4
-- 5~GHz. This frequency limit occurs due to ``temperature''
limitations posed to integration level and switching rate of
transistors. Note that cryogenic cooling of semiconductor chips will
not solve the problem \cite{8,9}.

Future of high performance computing is most likely associated with
one of alternative ``Post-Moore's'' technologies where energy
dissipation is drastically lower. It is expected that leader will be
determined by 2030, while 2020 -- 2030 will be the ``decade of
diversity''. In this paper, we consider one of the most promising
candidates for leadership -- the superconductor digital technology.
Basic element switching energy here is of the order of $10^{-19}$~J,
with no penalty for signal transfer. For a certain algorithm
superconducting circuits were shown to be up to seven orders of
magnitude more energy efficient than their semiconductor
counterparts, including power required for cryogenic cooling
\cite{AQFPrec}. Maturity level of superconductor technology can be
illustrated by the notional prototype of the superconducting
computer being developed under IARPA programm ``Cryogenic computing
complexity'' \cite{C3}. This is a 64~bit computing machine operating
at 10~GHz clock frequency with throughput of $10^{13}$~bit-op/s and
energy efficiency of $10^{15}$~bit-op/J at 4~K temperature.
Prospective study shows that superconductor computer could
outperform its semiconductor counterparts by two orders of magnitude
in energy efficiency, showing $250$~Gflops/W \cite{Holmes}.

Purpose of this paper is a review of superconducting logic and
memory circuits principles of operation and analysis of their issues
in respect to computer circuits design. We certainly do not claim to
be comprehensive while considering only the most common solutions.
Our review contains two main parts describing logic and memory,
correspondingly.

In the first part we start with examination of physical basis
underlying logic circuits operation. Superconductor logics are
presented by two main branches: digital single flux quantum (SFQ)
logics and adiabatic superconductor logic (ASL). Basic principles of
SFQ circuits operation are shown on example of the most popular
rapid single flux quantum (RSFQ) logic. It's energy efficient
successors and competitor, LV-RSFQ, ERSFQ, eSFQ and reciprocal
quantum logic (RQL), are considered subsequently. ASL is described
in the historical context of its development for ultra energy
efficient reversible computation. The modern status is presented by
two implementations of this logic. It can be noted that
superconducting adiabatic cells are used also in quantum computer
circuits like the ones fabricated by D-Wave Systems.

The second part of the review is devoted to cryogenic memory. Four
approaches are described: SQUID-based memory, hybrid Josephson-CMOS
memory, Josephson magnetic random access memory (JMRAM), and
orthogonal spin transfer magnetic random access memory (OST-MRAM).
They are presented in historical order of their development. In the
end of each part of our review we briefly discuss major challenges
and directions of possible further research in the studied area.

\section*{Review}

\section{Logic}

\subsection{Physical basis underlying logic circuits}

The fundamental physical phenomena underlying superconducting logic
circuits operation is the superconductivity effect, the quantization
of magnetic flux and the Josephson effects. The first one enables
ballistic signal transfer not limited by power necessary to charge
capacitance of interconnect lines. It provides the biggest advantage
in energy efficiency in comparison to conventional CMOS technology.
Indeed, superconducting microstrip lines are able to transfer
picosecond waveforms without distortions with speed approaching the
speed of light, for distances well exceeding typical chip size, and
with low crosstalk \cite{LSmsl}. This is the basis for fast
long-range interactions in superconducting circuits.

Note that absence of resistance ($R = 0$) leads to absence of
voltage ($V = 0$) in a superconducting circuit in stationary state.
Superconducting current flow corresponds not to electrical
potentials difference (the voltage $V = \delta \phi$) but to
difference of superconducting order parameter phases $\delta\theta$,
accordingly. Superconducting order parameter corresponds to
superconducting electrons wave function $|\psi|e^{i\theta}$ in
Ginzburg -- Landau theory \cite{GL}. Magnetic flux $\Phi$ in a
superconducting loop of inductance $L$ provides an increase of
superconducting phase along the loop and results in permanent
circulating current $I = \Phi/L$. This ratio is analogous to Ohm's
law $I = V/R$. It allows to write linear Kirchoff equations for
superconducting circuits.

The quantization of magnetic flux introduces fundamental difference
between CMOS and superconducting circuits operation. It follows from
uniqueness of superconducting electrons wave function. Indeed,
increase of superconducting phase along a loop corresponds to
magnetic flux as $\Phi = (\Phi_0/2\pi) \oint \nabla \theta dl$
(where $\Phi_0 = h/2e \approx 2 \times 10^{-15}$~Wb is the magnetic
flux quantum, $h$ is the Planck constant, and $e$ is the electron
charge). Fulfillment of this relation is possible if $\oint \nabla
\theta dl = 2\pi n$ (where $n$ is integer) and therefore $\Phi =
n\Phi_0$. Magnetic flux in a superconducting loop can take only
values multiple to the flux quantum, accordingly.

Physical representation of information is typically based on the
quantization of magnetic flux. For example, presence or absence of
SFQ in a superconducting loop can be considered as a logical unity
``1'' or zero ``0''. Note that information is physically localized
due to such representation. This is a fundamental difference
compared to information representation in semiconductor circuits.
The localization leads to deep analogy between superconducting logic
cells and von Neuman cellular automata \cite{LSmsl} where
short-range interactions are predominant.

The nonlinear element in superconducting circuits is the Josephson
junction. It is a weak link between two superconductors, e.g., the
most used superconductor-isolator-superconductor (SIS) sandwich. One
of the most important Josephson junction parameters is its critical
current, $I_c$. This is the maximum superconducting current capable
of flowing through the junction. Josephson junction can be switched
from superconducting to resistive state by increasing the current
above $I_c$. Transition to resistive state allows to change magnetic
flux in a superconducting loop, and hence to perform a digital logic
operation.

Dynamics of SIS junction is commonly described in the frame of the
resistively shunted junction model with capacitance (RSJC)
\cite{RSJC}. This model presents Josephson junction as a parallel
connection of the junction itself transmitting superconducting
current, $I_s$, only, a resistor and a capacitor with corresponding
currents, $I_r = V/R$ and $I_{cap} = C(\partial V/\partial t)$,
where $t$ is time. The total current through the junction is the
sum, $I = I_s + I_r + I_{cap}$. This model is based on DC and AC
Josephson effects which determine the superconducting current $I_s$
and voltage $V$.

The DC Josephson effect describes the superconducting current-phase
relation (CPR). For SIS junction it is $I_s = I_c\sin\varphi$, where
$\varphi = \nabla\theta$ is the superconducting order parameter
phase difference across the Josephson junction. It is called the
Josephson phase. By presenting the relation between superconducting
order parameter phase and magnetic flux as $\varphi = 2\pi
\Phi/\Phi_0$, we note that CPR couples current with the magnetic
flux in a superconducting loop. Josephson junction acts as a
nonlinear inductance in the circuits, accordingly.

The AC Josephson effect binds the voltage on Josephson junction in
resistive state with the superconducting phase evolution as $V =
(\Phi_0/2\pi)[\partial\varphi/\partial t]$. According to this
relation, increase of the Josephson phase in $2\pi$ is accompanied
by appearance of the voltage pulse across the junction such that
$\int Vdt = \Phi_0$. Therefore, a single switching of the Josephson
junction into resistive state corresponds to transmission of SFQ
pulse through the junction. The energy dissipated in the switching
process is $E_J \approx I_c \Phi_0 \approx 2 \times 10^{-19}$~J,
taking typical $I_c \approx 0.1$~mA. The typical critical current
value is conditioned by working (liquid helium) temperature, $T =
4.2$~K. For proper circuits operation it should be about three
orders higher than the effective noise current value, $I_T =
(2\pi/\Phi_0) k_B T \approx 0.18$~$\mu$A, where $k_B$ is the
Boltzmann constant.

Characteristic frequency of the Josephson junction switching process
is determined by Josephson junction parameters, $\omega_c =
(2\pi/\Phi_0) I_c R_n$, where $I_c R_n$ product is the Josephson
junction characteristic voltage, and $R_n$ is the junction
resistance in the normal state. Since SIS junctions possess large
capacitance, they are usually shunted by external resistors to avoid
$LC$-resonances. The resistance $R_n$ is approximately equal to
resistance of the shunt $R_n \approx R_s$ because $R_s$ is much
smaller than own tunnel junction resistance. For Nb-based junctions
the characteristic frequency is of the order of $\omega_c/2\pi \sim
100 - 350$~GHz (the characteristic voltage is at the level of $\sim
0.2 - 0.7$~mV). Superconducting digital circuits are predominantly
based on tunnel junctions because of high accuracy of their
fabrication process and high characteristic frequencies.

By expressing the currents $I_s$, $I_r$ and $I_{cap}$ of RSJC model
through the Josephson phase $\varphi$, we can present the total
current flowing through the junction in the following form:
\begin{equation} \label{RSJC}
I/I_c = \sin\varphi + \omega_c^{-1}\dot{\varphi} +
\beta_c\omega_c^{-2}\ddot{\varphi},
\end{equation}
where $\beta_c = \omega_c R_n C$ is the Stewart-McCumber parameter
reflecting capacitance impact, and dot denotes time differentiation.
This Equation \ref{RSJC} is quite analogous to the one for
mechanical pendulum with the moment of inertia $\beta_c/\omega_c^2$
(capacitance here is analogous to mass), the viscosity factor
$1/\omega_c$ (resistance determines damping), and the applied torque
$I/I_c$. This simple analogy allows to consider superconducting
digital circuit as a net of coupled pendulums.

\begin{figure}
\includegraphics[bb = 0 0 205 108,width=0.8\columnwidth,keepaspectratio]{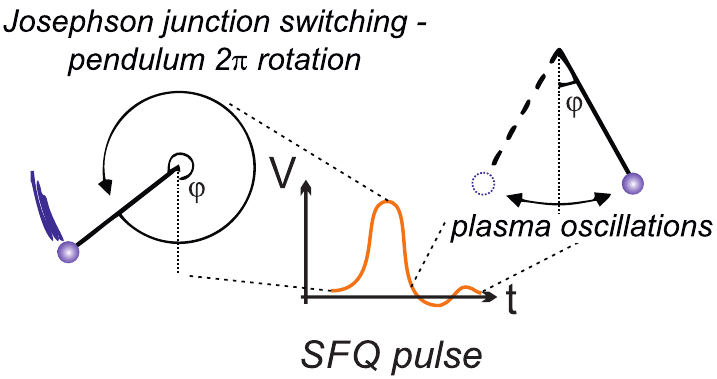}
\caption{Voltage pulse on Josephson junction corresponding to SFQ
transition and its mechanical analogy with pendulum rotation.}
\label{Fig1}
\end{figure}

Pendulum $2\pi$ rotation is accompanied by subsequent oscillations
around stable equilibrium point (Figure~\ref{Fig1}). In Josephson
junction dynamics they are called ``plasma oscillations''. Plasma
oscillations frequency is $\omega_p = \omega_c/\sqrt{\beta_c} =
\sqrt{2\pi I_c/\Phi_0 C}$. For proper logic cell operation these
oscillations should vanish before subsequent Josephson junction
switching. Compliance with this requirement can be achieved with
$\beta_c \approx 1$, $\omega_p \approx \omega_c$. Clock frequency is
accordingly less than $\omega_c$, and is under $100$~GHz in
practical circuits.

Complexity of a superconducting circuit realizable on a chip is
determined by Josephson junction dimensions. Area of Josephson
junction is closely related to its critical current density, $j_c$.
This parameter is one of the most important in the standard Nb-based
tunnel junction fabrication process. It is fixed by material
properties of insulating interlayer Al$_2$O$_3$ between
superconducting Nb electrodes, and its thickness $d \approx 1$~nm.
The critical current density value lies typically in the range $j_c
= 10 - 100$~$\mu$A/$\mu$m$^2$. The corresponding Josephson junction
specific capacitance is $c \approx 40 - 60$~fF/$\mu$m$^2$. Variation
in Josephson junction critical current, $I_c = a j_c$, is obtained
by variation of its area, $a$. It is accompanied by variation of
Josephson junction capacitance, $C = a c$. The shunt resistance is
adjusted in accordance with the mentioned condition, $\beta_c = 1$,
as $R_n = \sqrt{\Phi_0/2\pi j_c c}/a$. Its area is defined by
Josephson junction area $a$, minimum wiring feature size
\cite{T,HYPDL} ($\sim 0.5 - 1$~$\mu$m), and sheet resistance of used
material ($2 - 6$~$\Omega$ per square for Mo or MoN$_x$)
\cite{T,HYPDL}.

While own Josephson junction weak link area is typically $a \sim
1$~$\mu$m$^2$ for $j_c = 100$~$\mu$A/$\mu$m$^2$, its total area with
the shunt is by an order lager. Corresponding Josephson junctions
available density on a chip is $10^7/$cm$^2$. Superconducting
circuit complexity becomes limited to $2.5$ million junctions per
$1$~cm$^2$ nowadays under assumption that only a quarter of chip
area can be occupied by Josephson junctions (with taking
interconnects into account) \cite{T}. The circuit can be further
expanded using multi-chip module (MCM) technology \cite{MCM1,MCM2}.

\subsection{Digital SFQ logics}

\subsubsection{SFQ circuit basic principles of operation}

\begin{figure}
\includegraphics[bb = 0 0 180 72,width=0.7\columnwidth,keepaspectratio]{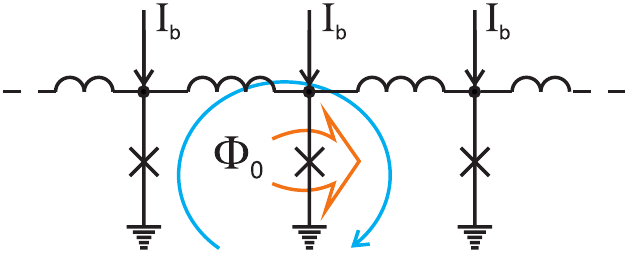} \caption{Josephson
transmission line. Josephson junctions are shown by crosses. $I_b$
is applied bias current. Blue arrow presents SFQ circulating
current. Orange arrow highlights Josephson junction switched in
resistive state.} \label{Fig2}
\end{figure}

Data processing in SFQ circuits can be discussed on an example of
RSFQ cells operation. RSFQ data bus is shown in Figure~\ref{Fig2}.
It is a parallel array of superconducting loops composed of
Josephson junctions (shown by crosses) and superconducting
inductances. This structure is called the Josephson transmission
line (JTL). SFQ can be transferred along this JTL by successive
switchings of Josephson junctions. The switching is obtained by
summing the SFQ circulating current and the applied bias current
$I_b$. Josephson junction transition into resistive state leads to
SFQ circulating current redistribution toward the next junction. The
redistribution process ends by the next junction switching and
successive returning of the current junction into the
superconducting state.

This example shows the basic principle of SFQ logic cells operation.
It reduces to summation of currents, which are SFQs currents and
bias currents. This summation leads (or not leads) to successive
Josephson junction switching resulting in reproduction (or not
reproduction) of SFQ. In RSFQ convention \cite{LSmsl,20} arrival of
an SFQ pulse during clock period to a logic cell has a meaning of
binary ``1'', while absence of the one means ``0''.

\begin{figure}
\includegraphics[bb = 0 0 183 118,width=0.75\columnwidth,keepaspectratio]{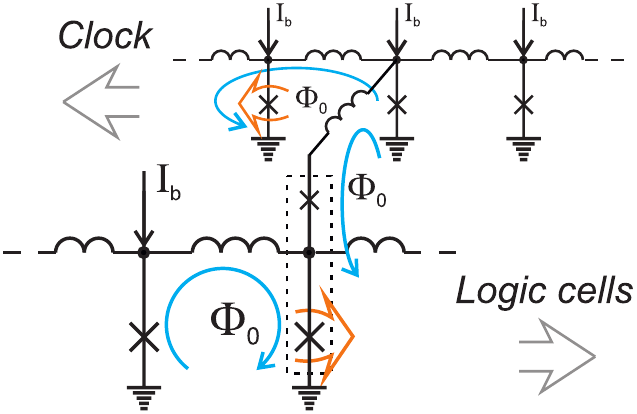} \caption{RSFQ
logic cell coupled to clocking JTL. $I_b$ is applied bias current.
Blue arrows present circulating currents of SFQs. Orange arrows
highlight Josephson junctions switched in resistive state. Dotted
rectangle marks decision making pair.} \label{Fig3}
\end{figure}

Figure~\ref{Fig3} illustrates an example of clocked readout of
information from a RSFQ logic cell. Clocking is performed by means
of SFQs application to the cell. Upper JTL in Figure~\ref{Fig3}
serves for SFQ clock distribution. SFQs are allotted to the cell
through extra branch coupled to the JTL as shown. Note that
Josephson junction clones SFQ at the branch point. Readout operation
is performed by a couple of junctions marked by dotted rectangle.
This couple is commonly called the decision making pair. Existence
(or absence) of an SFQ circulating current in the logic cell loop
makes the lower junction to be closer (or father) to its critical
current compared to the upper junction. Clocking SFQ switches the
lower (or upper) junction, correspondingly. SFQ reproduction by the
lower junction means logical ``1'' to the output, while SFQ absence
from the ones means logical ``0''.

One can note a couple of typical SFQ circuits features from the
presented example. Considered logic cell acts as a finite state
machine. Its output depends on a history of its input. This
particular cell operates as a widely used D flip-flop (``D'' means
``data'' or ``delay'') -- the basis of shift registers. Note that
its realization is much simpler than the one of semiconductor
counterparts. RSFQ basic cells are such flip-flops, and therefore
RSFQ is sequential logic. This is in contrast with semiconductor
logic which is combinational one (where logic cell output is a
function of its present input only).

Since only one clocked operation is performed during a clock period
(some operations can be performed asynchronously), a processing
stage in RSFQ circuits is reduced to a few logic cells. This is also
completely opposite to conventional semiconductor circuits.

\subsubsection{RSFQ logic}

RSFQ logic dominates in superconductor digital technology since
1990-s years \cite{17}. Many digital and mixed signal devices like
analog-to-digital converters \cite{OM_ADC,OM_ADC1}, digital signal
and data processors \cite{M_DCRev} were realized on its basis.

Unfortunately, energy efficiency did not matter in the days of RSFQ
development. High clock frequency was thought to be the major RSFQ
advantage in the beginning. Extremely fast RSFQ-based digital
frequency divider \cite{18} (T flip-flop) was presented just about a
decade later RSFQ invention. Its clock frequency was as high as
$770$~GHz. It is still among fastest ever digital circuits.

\begin{figure}
\includegraphics[bb=0 0 138 84,width=0.6\columnwidth,keepaspectratio]{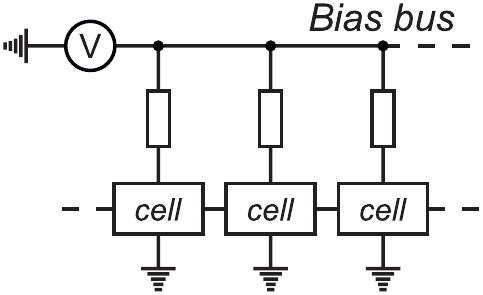} \caption{RSFQ
power supply scheme.} \label{Fig4}
\end{figure}

The first RSFQ basic cells were the superconducting loops with two
Josephson junctions (commonly known as the superconducting quantum
interference devices - SQUIDs). These cells were connected by
resistors \cite{20,19} (so ``R'' was for ``resistive'' in the
abbreviation). Power supply bus coupling was also resistive. While
resistors connecting the cells were rather quickly substituted for
superconducting inductances and Josephson junctions \cite{RSFQind},
the ones in feed lines remained until recent years, see
Figure~\ref{Fig4}. They determined stationary energy dissipation,
$P_S = I_b V_b$, where $I_b$ and $V_b$ are the DC bias current and
according voltage. The bias current is typically $I_b \approx 0.75
I_c$. The bias voltage had to be an order higher than the Josephson
junction characteristic voltage, $V_b \sim 10 \times I_c R_n$, to
prevent the bias current redistribution. This requirement determined
the bias resistors values. Typical RSFQ cell stationary power
dissipation \cite{8} is $P_S \sim 800$~nW.

Another mechanism providing power dissipation corresponds to
Josephson junction switching. This dynamic power dissipation is
defined as $P_D = I_b\Phi_0 f$, where $f$ is the clock frequency.
For a typical clock frequency of 20 GHz $P_D$ is at the level
\cite{8} of $\sim 13$~nW. It is seen that the dynamic power
dissipation is about 60 times less than the stationary one. Main
efforts to increase RSFQ circuits energy efficiency were aimed at
stationary energy dissipation decrease, accordingly. RSFQ energy
efficient successors, LV-RSFQ, ERSFQ and eSFQ, are presented below.

\subsubsection{LV-RSFQ}

The first step toward $P_S$ reduction was the bias voltage decrease.
Bias current redistribution between neighboring cells in low-voltage
RSFQ (LV-RSFQ) is damped by introduction of inductances connected in
series with bias resistors in feed lines \cite{23,24,25,26,261}.

Unfortunately, this approach limits clock frequency. Indeed, clock
frequency increase is accompanied by increase of average voltage
$\overline{V}$ across a cell (according to the AC Josephson effect).
This in turn leads to bias current decrease proportional to $V_b -
\overline{V}$. The latter finally results in the cell malfunction
\cite{RL}. This tradeoff with requirement of additional circuit area
for inductances in feed lines practically limit application of this
approach. Since static power dissipation is not eliminated, this is
somewhat half-hearted solution. It was succeeded by another two RSFQ
versions (ERSFQ and eSFQ, where ``E/e'' stays for ``energy
efficient'') where $P_S$ is totaly zero.

\subsubsection{ERSFQ}

\begin{figure}
\includegraphics[bb=0 0 121 108,width=0.55\columnwidth,keepaspectratio]{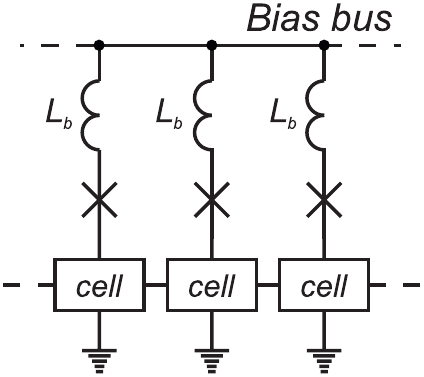} \caption{ERSFQ
power supply scheme. $L_b$ is inductance limiting bias current
variation.} \label{Fig5}
\end{figure}

ERSFQ \cite{27} is the next logical step after LV-RSFQ. Resistors in
feed lines are substituted for Josephson junctions limiting bias
current variation in this logic, see Figure~\ref{Fig5}. This
replacement is somewhat analogous to the one which was done for
resistors connecting SQUID cells in the very first RSFQ circuits. It
provides possibility for the circuits to be in pure superconducting
state.

Main difficulty in the bias resistors elimination is formation of
superconducting loops between logic cells. Generally, logic cells
are switched asynchronously depending on processing data. Average
voltage and total Josephson phase increment are different across
them. This results in emergence of currents circulating through
neighbor cells. Being added to bias current, these currents prevent
correct operation of the circuits.

Imbalance of Josephson phase increment is automatically compensated
by corresponding switchings of Josephson junctions placed in ERSFQ
feed lines. Since these switchings are not synchronized with clock,
some immediate alteration of bias current is still possible. This
alteration $\Delta I \sim \Phi_0/L_b$ is limited by inductance $L_b$
connected in series with Josephson junction in the feed line. While
large value of this inductance $L_b$ minimizes the bias current
variation, its large geometric size increases the circuit area
(similar to LV-RSFQ). Possible solutions of this problem are an
increase of wiring layers number, and utilization of superconducting
materials having high kinetic inductance. These materials can be
also used for further miniaturization of logic cells themselves
\cite{T}.

\subsubsection{eSFQ}

\begin{figure}
\includegraphics[bb=0 0 165 102,width=0.6\columnwidth,keepaspectratio]{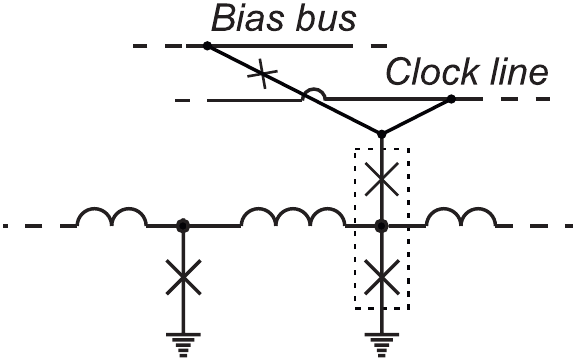} \caption{eSFQ
power supply scheme. Dotted rectangle marks decision making pair.}
\label{Fig6}
\end{figure}

Another energy efficient logic of RSFQ family is eSFQ
\cite{8,28,32,33}. The main idea here is the ``synchronous phase
balancing''. Bias current is applied to decision making pair, see
Figure~\ref{Fig6}. One Josephson junction of this pair is always
switched during a clock cycle regardless data content. Therefore,
average voltage and Josephson phase increment are always equal
across any such pair. This prevents the emergence of parasitic
circulating currents. Josephson junction in the feed line is
required only for proper phase balance adjustment during power-up
procedure. ``It is not expected to switch during regular circuit
operation'' \cite{8}.

Achieved phase balance allows to eliminate large inductances from
ERSFQ feed lines, and so eSFQ circuits occupy nearly the same area
as RSFQ ones. One should note that despite of the ``synchronous''
nature of this logic, a method for design of eSFQ-based asynchronous
circuits was proposed in \cite{32}, making it suitable for
wave-pipelined architecture.

Since RSFQ library was designed regardless synchronous phase
balancing, transition to eSFQ requires its correction. In some cases
it leads to increase of Josephson junctions number. For example, JTL
should be replaced by a shift register \cite{29} or by ``Wave JTL''
\cite{32}, or by one of its asynchronous counterparts: ballistic
transmission line based on unshunted Josephson junctions
\cite{30,31} or passive microstrip line.

Similarity of ERSFQ and eSFQ approaches allows to make an overall
assessment of total increase in Josephson junctions number up to
$33-40$\% compared to RSFQ circuits \cite{8}. Inheritance of basic
cells design of RSFQ by ERSFQ makes it easier to use.

\subsubsection{RSFQ logic family common features}

Clock is effectively a part of data in ERSFQ circuits. This means
that they are globally asynchronous.

Since clock frequency is determined by repetition rate of SFQs in
clocking JTL, it can be adjusted ``in flight'' by logic cells
according to processing data.

\begin{figure}
\includegraphics[bb=0 0 331 112,width=1\columnwidth,keepaspectratio]{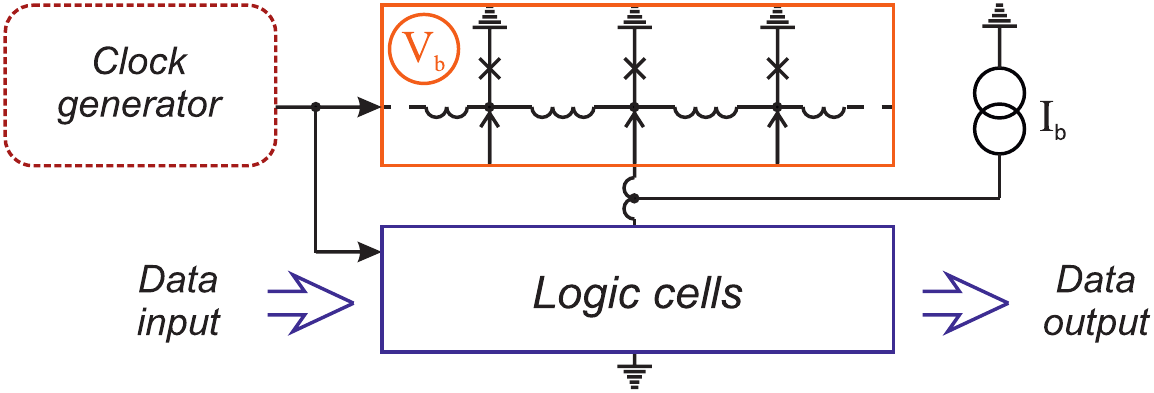} \caption{DC bias
voltage source realization in RSFQ circuitry.} \label{Fig7}
\end{figure}

The bias voltage source can be implemented as a JTL fed by a
constant bias current, for which the input signal is the SFQ clock
applied from an on-chip SFQ clock generator, see Figure~\ref{Fig7}.
Average voltage on this JTL is precisely proportional to the clock
frequency, $\overline{V}_b = \Phi_0 f$, according to the AC
Josephson effect. Clock control by logic cells allows to adjust this
voltage or even to turn it off. The last option corresponds to
circuits switching into ``sleep mode'' where power dissipation is
totally zero. Realization of this power save mechanism at individual
circuits level is possible with circuits partitioning into series
connection of islands with equal bias current but different bias
voltage \cite{T_RSFQ_SB}.

Since logic cells are fed in parallel in RSFQ logic family, total
bias current increases proportional to Josephson junctions number.
For 1 million Josephson junctions the bias current value could be
unreasonably high $I_b \sim 100$~A. Circuits partitioning allows to
keep it at acceptable level \cite{Jap_HS_SFQ} below $3$~A.

\subsubsection{RQL}

\begin{figure}
\includegraphics[bb=0 0 131 93,width=0.55\columnwidth,keepaspectratio]{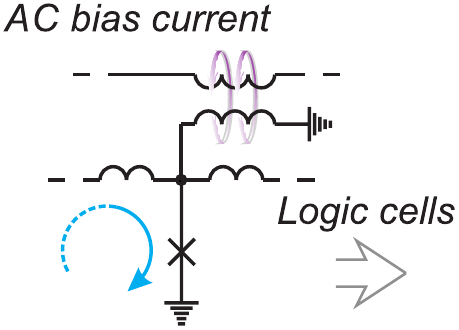} \caption{RQL AC
power supply scheme. Blue arrow shows SFQ current, violet arrows
present magnetic coupling.} \label{Fig8}
\end{figure}

RQL was proposed in about 2008. It was developed as an alternative
to conventional RSFQ, and presented as ``ultra-low-power
superconductor logic'' \cite{14}. Main difference between RQL and
RSFQ is in the way of power supply \cite{RQL}. While in RSFQ it is a
DC power applied to Josephson junctions in parallel through bias
resistors (Figure~\ref{Fig4}), in RQL it is an AC power applied in
series through bias transformers, see Figure~\ref{Fig8}.

The proposed power supply scheme possesses some advantages. (i) No
DC bias current and no bias resistors means zero static power
dissipation inside cryogenic cooler. Bias current is terminated off
chip at room temperature. (ii) The well known RSFQ circuits design
problem is the large return bias current magnetic field affecting
logic cells. It is recommended \cite{Jap_HS_SFQ} to keep maximum
bias current below $100$~mA in RSFQ feed line. This return current
is completely absent in RQL due to the mentioned off-chip bias
current termination. (iii) Serial bias supply allows to keep bias
current amplitude at fairy low level \cite{14} of the order of $I_b
\sim 1.8$~mA regardless number of Josephson junctions on a chip.
There is no need for the large-scale circuit partitioning. (iv) Bias
current plays a role of clock signal. There is no need for SFQ clock
distribution network. (v) Clock is not affected by thermal noise.

\begin{figure}
\includegraphics[bb=0 0 469 173,width=1\columnwidth,keepaspectratio]{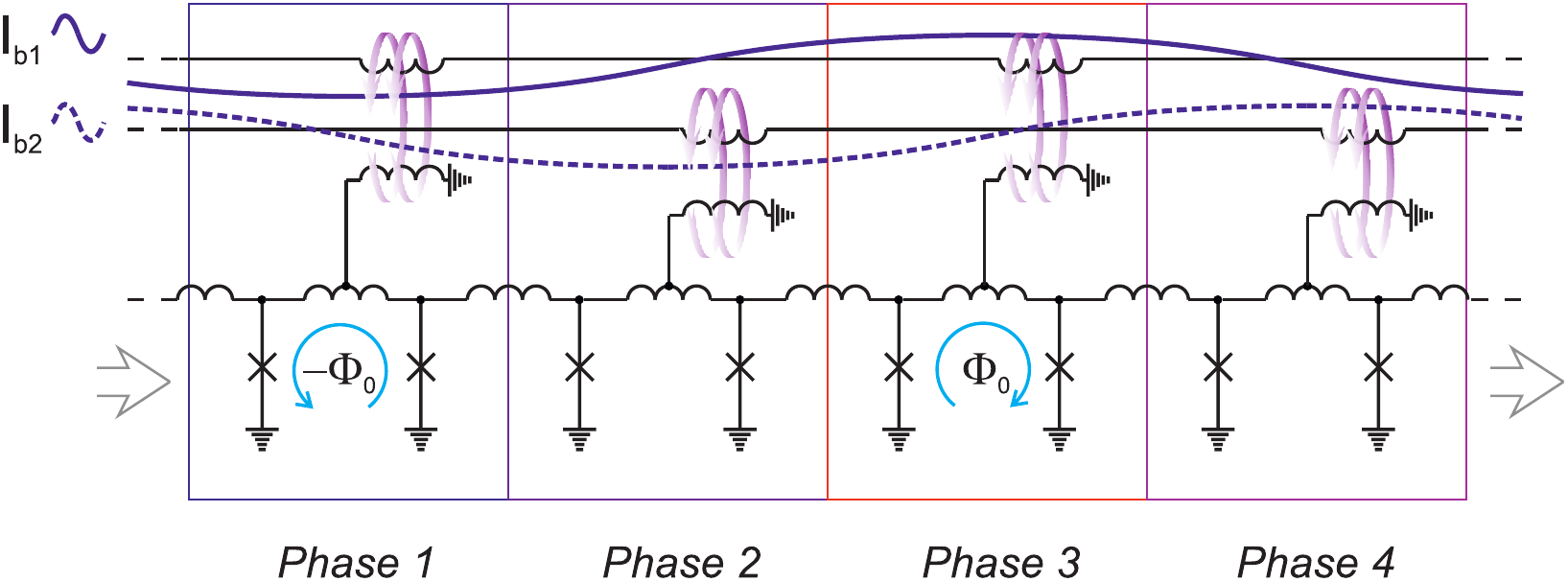} \caption{RQL
transmission line with four-phase bias. $I_{b1,2}$ are AC bias
currents providing power supply, and playing role of clock signal.
Blue arrows show SFQ currents, violet arrows present magnetic
coupling.} \label{Fig9}
\end{figure}

Logical unity (zero) is presented by a pair of SFQs having opposite
magnetic flux directions (or lack thereof) in RQL circuits. These
SFQs can be transferred in one direction with application of
inversely directed bias currents, see Figure~\ref{Fig9}. The SFQs
are placed in positive/negative AC current wave half period,
accordingly. Unfortunately, one AC bias current is insufficient for
directional propagation of the SFQs. It can provide only periodic
space oscillations of the flux quanta. RQL uses two AC bias currents
with $\pi/2$ phase shift. RQL cells are coupled to these two feed
lines in rotation (Figure~\ref{Fig9}). Such coupling produces space
division of total bias current/clock into four windows shifted by
$0$, $\pi/2$, $\pi$, and $3\pi/2$ wave period. By analogy with a
car's four-stroke engine, this four-phase bias scheme provides
directionality of the SFQs propagation \cite{14}.

Logic elements connected to a single AC bias line within a single
clock phase window form a pipeline. The pipeline in RQL can contain
an arbitrary number of cells. One can increase a depth of the
pipeline at the cost of clock frequency decrease. Time delay of the
pipeline should be less than one-third of clock period for proper
circuit operation. Circuit speed is effectively a product of clock
frequency and pipeline depth. Maximum clock frequency of RQL
circuits can be estimated as $f_{max} \sim 17$~GHz under assumption
of the standard Josephson junction characteristic frequency
$\omega_c/2\pi = 350$~GHz and $N = 8$ Josephson junctions in the
pipeline \cite{RQL}.

RQL biasing scheme provides self data synchronization. Early pulses
wait at the pipeline edge for bias current rise in the next phase
window. SFQ jitter is accumulated only inside one pipeline, and
therefore timing errors are negligible which is in contrast to RSFQ.

RQL logic cells are state machines similar to RSFQ ones. Internal
state of logic cell can be changed by SFQ propagating in front of
the clock wave. Its paired SFQ with opposite polarity serves for the
state resetting in the end of a clock period.

Complete set of RQL logic cells comprises just three gates which are
And-Or gate, A-not-B gate and Set-Reset latch. These gates behave as
combinational logic cells similar to their semiconductor
counterparts \cite{RQL}. This makes RQL circuits design to be closer
to CMOS than to RSFQ.

Particular RQL drawbacks come from power supply scheme as well as
its advantages. Proper power supply requires high-frequency power
splitters. These splitters often occupy quite a large area. For
example, in implementation of 8-bit carry-look-ahead adder they
cover area $\sim 2.5$ times larger than the adder itself
\cite{RQL_adder}. One can note that power supply through
transformers also limits the possibility for the circuits
miniaturization.

Multiphase AC bias presents known difficulty for high-frequency
design (clock skew etc.). This practically limits clock frequency to
$10$~GHz, while RSFQ circuits routinely operate at frequency of
$50$~GHz. Moreover, implementation of MCM technology becomes
complicated with RQL due to possible asynchronization of chips or
clock phase shift. Besides inconvenience presented by high-frequency
clock supply from off-chip external source, clocking by AC bias
currents eliminates possibility of clock control by logic cells.
Corresponding power save mechanisms cannot be realized in RQL. In
addition, one should mention RF losses in microstrip resonators
which typically make up to 50\% total power budget even at
relatively low frequencies.

Total power dissipation of RQL and ERSFQ circuits in active mode
seems similar. Static power dissipation is absent. Dynamic power
dissipation is associated with Josephson junction switching in data
propagation process. In RQL circuits Josephson junction is doubly
switched for logical unity transfer and zero times for transfer of
logical zero. In ERSFQ both unity and zero are transferred with
switching of one of the Josephson junctions in decision making pair.
By assuming equal number of zeros and ones in data, one comes to
roughly equal estimation for energy dissipation in both RQL and
ERSFQ logics \cite{T}. More detailed analysis shows that only
adiabatic switching of logic cells improves superconducting circuits
energy efficiency markedly \cite{T}.

\subsection{Adiabatic superconductor logic}

Considered variants of superconductor logics have been proposed for
non-adiabatic irreversible computation. Logical states are separated
here by energy barrier $E_w \sim 10^3 - 10^4~k_B T$ ensuring proper
circuit operation. Note that the energy barrier in semiconductor
circuits is two to three orders higher, $E_w \sim 10^6~k_B T$.
Minimal energy barrier corresponds to Landauer's ``thermodynamic
limit'', \cite{40} $E_{min} = k_B T \ln 2$. In this limit logic
states distinguishability becomes completely lost due to thermal
fluctuations \cite{9}.

Energy required to perform a non-adiabatic logic operation can be
estimated as the energy of transition between logical states
corresponding to $E_w$. In considered superconductor logics it is
the energy of Josephson junction switching, $E_J \approx 2 \times
10^{-19}$~J. While presuming logical irreversibility, this energy
can be lowered down to $E_{min} \approx 4 \times 10^{-23}$~J (at $T
= 4.2$~K) by using adiabatic switching process. Note that Landauer
limit $E_{min}$ in this context reflects computing system entropy
change associated with an irreversible operation \cite{40}. At the
same time, there is no such limit for physically and logically
reversible process. Therefore, energy dissipated per logical
operation can approach zero in adiabatic reversible circuits.

The first ever practical reversible logic gates were realized
recently \cite{48} on a basis of adiabatic superconductor logic.
History of ASL development have begun even before RSFQ invention
with proposition of ``parametric quantron'' \cite{PQ} in 1976. This
cell itself was proposed even earlier in 1954 as ``rf parametron''
\cite{rfP}, though for different operating regime.

It is interesting to note that the manner of parametric quantron
cell operation was implemented later in a single-electron device
\cite{17,QCA1} in 1996. The ``single-electron parametron'' operation
was in fact quite similar to the ones of quantum-dot cellular
automata (QCA) which were proposed for computation those years
\cite{QCA2}.

\subsubsection{ASL circuit basic principles of operation}

\begin{figure}
\includegraphics[bb=0 0 298 73,width=1\columnwidth,keepaspectratio]{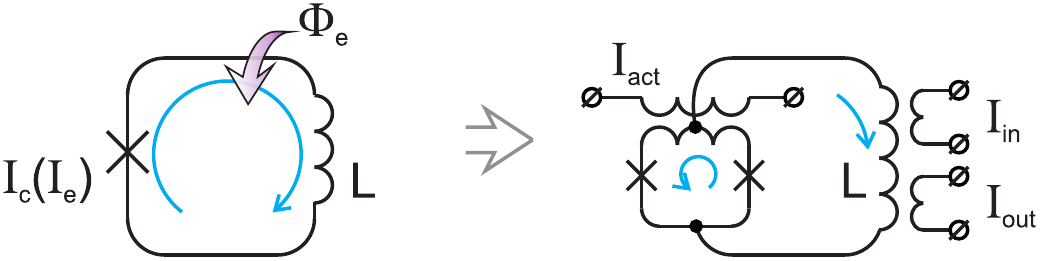}
\caption{Parametric quantron notional (left) and practical (right)
schematic. The cell state is conditioned by bias flux $\Phi_e$ and
current $I_e$ controlling Josephson junction critical current $I_c$.
$L$ is the loop inductance. In practice, single Josephson junction
is substituted by SQUID controlled by activation current $I_{act}$.
$I_{in}$/$I_{out}$ are input/output currents.} \label{Fig10}
\end{figure}

Parametric quantron is a superconducting loop with single Josephson
junction shown in Figure~\ref{Fig10} leftward. Its state is
conditioned by external magnetic flux, $\Phi_e$, and current, $I_e$,
controlling Josephson junction critical current $I_c(I_e)$.
Potential energy of this cell is a sum of Josephson junction energy,
$U_J = (E_J/2\pi)[1 - \cos\varphi]$ (followed directly from the DC
Josephson effect), and magnetic energy, $U_M = (E_J/2\pi)[\varphi -
\varphi_e]^2/2l$:
\begin{equation} \label{ParaQ}
U_{PQ} = \frac{E_J}{2\pi}\left[1 - \cos\varphi + \frac{(\varphi -
\varphi_e)^2}{2l}\right],
\end{equation}
where $\varphi_e = 2\pi\Phi_e/\Phi_0$, $l = 2\pi I_c L/\Phi_0$ is
the normalized loop inductance.

\begin{figure}
\includegraphics[bb=0 0 818 572,width=1\columnwidth,keepaspectratio]{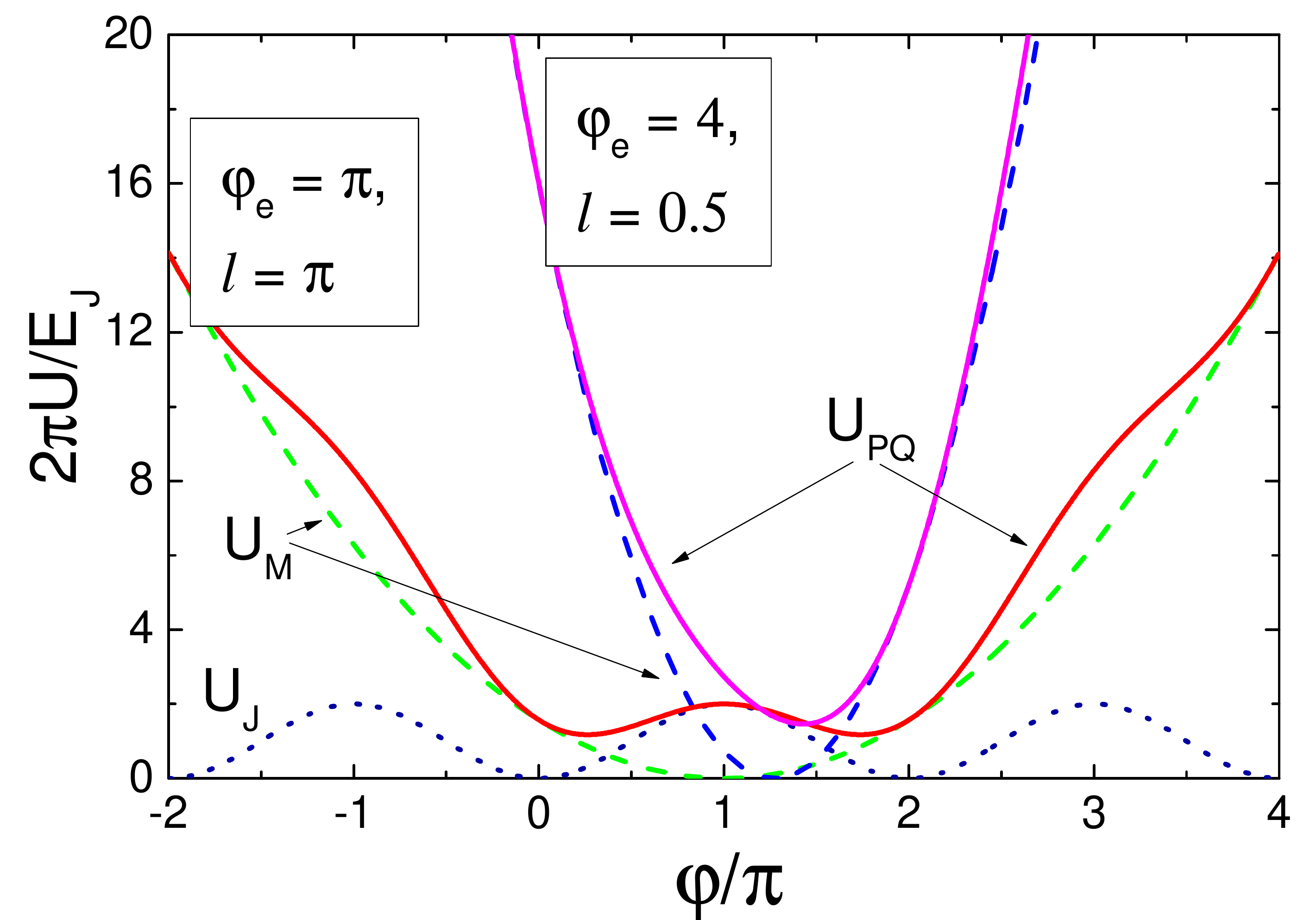}
\caption{Parametric quantron potential energy $U_{PQ}$ (\ref{ParaQ})
(solid lines) and its terms: magnetic energy $U_M$ (dashed lines),
and Josephson junction energy $U_J$ (dotted line).} \label{Fig11}
\end{figure}

It is seen that external parameters $\Phi_e$, $I_e$ control vertex
(through $\varphi_e[\Phi_e]$) and slope (through $l[I_c(I_e)]$) of
the potential energy parabolic term $U_M$ in Equation~\ref{ParaQ}.
Under appropriate bias flux, $\varphi_e \approx \pi$, parametric
quantron potential energy $U_{PQ}(\varphi)$ can take single-well (at
$l < 1$) or double-well (at $l > 1$) shape depending on $I_e$, see
Figure~\ref{Fig11}. Logical zero and unity can be represented by the
cell states with Josephson junction phase $\varphi$ lower or higher
than $\pi$. For $l > 1$ these states correspond to minima of
potential wells. Physically they correspond to different magnetic
flux in the loop (with current circulating in the loop in opposite
directions if $\varphi \neq 2\pi n$, where $n$ is integer).

\begin{figure}
\centering
\begin{minipage}{0.4\columnwidth}
\includegraphics[bb=0 0 177 141,width=1\columnwidth,keepaspectratio]{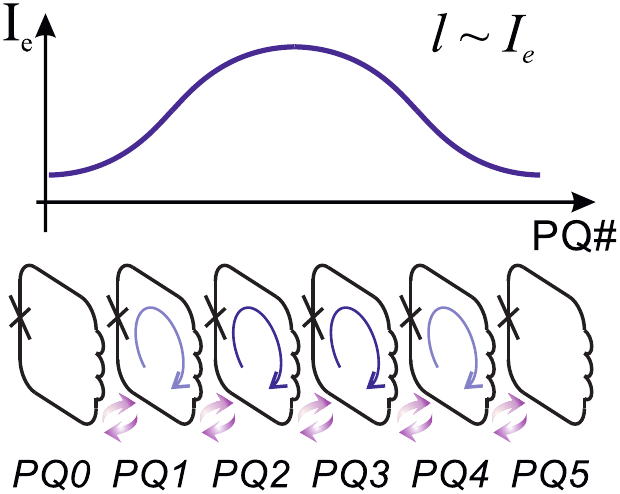}
\end{minipage}
\begin{minipage}{0.5\columnwidth}
\includegraphics[bb=0 0 818 572,width=1\columnwidth,keepaspectratio]{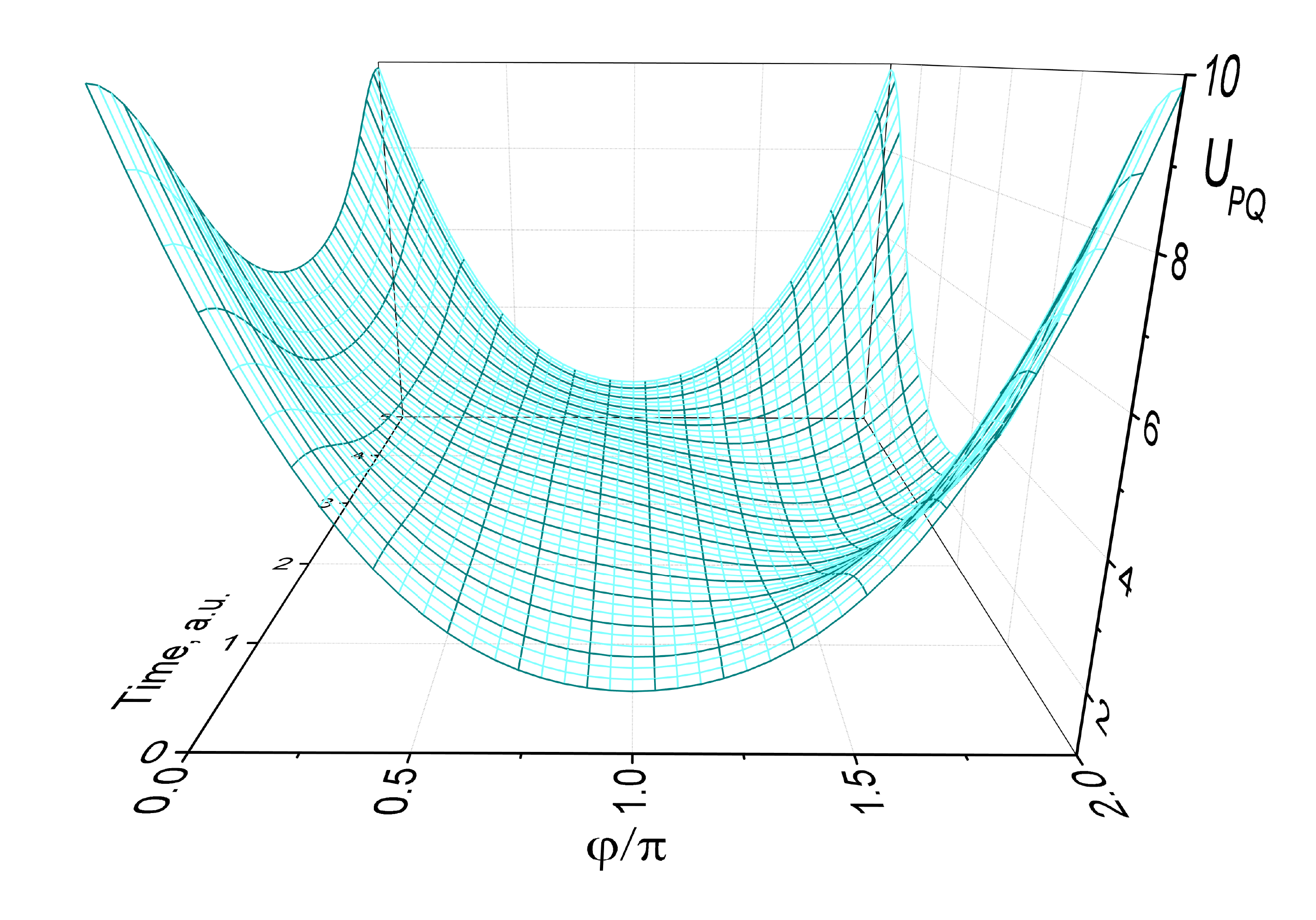}
\end{minipage}
\caption{Logical state transfer in array of magnetically coupled
parametric quantrons under driving current pulse $I_e$, with
corresponding change of a single cell potential profile in time.
Violet arrows present magnetic coupling.} \label{Fig12}
\end{figure}

Logical state transfer can be performed in array of magnetically
coupled parametric quantrons biased into working point $\varphi_e =
\pi$. Current pulse $I_e$ should be applied sequentially to the
cells increasing their normalized inductance one by one, see
Figure~\ref{Fig12}. Logical state can be shared by a group of cells,
wherein it is most pronounced in a cell with the largest $l$ in
particular moment. Dynamics of this transfer process can be made
adiabatic by adjusting the shape of the driving current pulse $I_e$.
Cross-coupling of the cells enables adiabatic reversible logic
operations \cite{LikhPQ}.

Single Josephson junction of parametric quantron was substituted by
SQUID (see Figure~\ref{Fig10} rightward) in practical
implementations \cite{Convolv}. Activation current, $I_{act}$, here
plays a role similar to the ones of $I_e$. It induces circulating
current in activation SQUID, and therefore increases Josephson
junctions phases, according to the DC Josephson effect. This in turn
corresponds to increase of Josephson junctions energy which can be
minimized with appearance of current circulating in the main
parametric quantron loop (the loop containing inductance $L$).
However, the states with both directions of the circulating current
are equally favorable due to symmetry of the scheme. Choice of one
of these states corresponds to direction of input current $I_{in}$
playing here the role of $\Phi_e$. Due to the fact that current-less
state is unstable balance corresponding to potential energy local
maximum, the current $I_{in}$ can be infinitesimally small.
Parametric quantron can provide virtually infinite amplification of
magnetic flux, accordingly. Since potential energy minimum is
achieved with circulating currents in both: activation SQUID and
main parametric quantron loop, it was noted that the roles of the
activation and the input/output can be swapped \cite{QFP}.

Unfortunately, already the first designs in mid 1980-s of physically
and logically reversible parametric quantron based processor
\cite{Convolv} showed this approach to be impractical. The reason
for such conclusion was as follows. Logical reversibility can be
achieved by temporary storage of all intermediate results
\cite{BenLR}. Together with predominance of short-range interactions
this produces severe hardware overhead. Indeed, realization of 8-bit
1024-points fast convolver required almost $10^7$ parametric
quantrons \cite{Convolv}. About 90\% of them were operated just as
elements of shift registers, transferring data through the processor
\cite{Convolv}. It was noted that such circuits are also featured by
low speed (in comparison to RSFQ) and low tolerance to parameters
variations \cite{17}.

\subsubsection{Quantum flux parametron based circuits}

Few years later the works on reversible circuits, the same
principles of operation were utilized for development of generally
non-reversible Josephson supercomputer. In this effort parametric
quantron was renamed as ``quantum flux parametron'' (QFP)
\cite{QFP}. The major problem of QFP-based circuits was
high-frequency multi-phase AC power supply (which was later borrowed
by RQL). While there were different approaches elaborated for its
solution \cite{QFP,QFP1}, finally multi-phase AC biasing was
recognized to be intractable obstacle for implementation of complex
high-clock-frequency practical circuits and QFP-based approach was
abandoned for some years.

Renewed interest to ASL was introduced by development of
superconductor quantum computer. QFPs are utilized as qubits and
couplers in adiabatic quantum optimization systems of D-Wave Systems
\cite{DWave1,DWave2}. Another reason for the current rise of
interest to ASL is Japan JST-ALCA project ``Superconductor
electronics system combined with optics and spintronics''
\cite{ALCA}. The idea of the project is the development of energy
efficient supercomputer based on synergy of the technologies.
Superconductor processor of the computing system is planned to be
based on QFPs operated in adiabatic regime. The processor prototype
has 8-bit simplified RISC architecture and is featured by $\sim 25$
thousand Josephson junctions and $\sim 10$ instructions. In this
context, adiabatic operation of QFP was investigated in order to
reduce its dynamic energy consumption down to the fundamental limit
\cite{45}. Adiabatic QFP was abbreviated as AQFP in these works
\cite{48,45,47,46,AQFPlatch,49,50}.

AQFP-based circuits were tested experimentally \cite{47} at $5$~GHz
clock frequency showing energy dissipation at the level of
$10^{-20}$~J. Theoretical analysis reveals that AQFP can be operated
with energy dissipation less than the thermodynamic limit \cite{46}.
Product of energy dissipated per clock cycle on a cycle time could
approach the quantum limit \cite{50} at $4.2$~K cooling temperature,
with utilization of standard manufacturing processes \cite{54}.
Comparison of AQFP-based design with design based on CMOS FPGA, on
example of implementation of Collatz algorithm, showed that the
first one is about seven orders of magnitude superior to its
counterpart in energy efficiency, even including the power of
cryogenic cooling \cite{AQFPrec}.

AQFP-based logic cells can be implemented by combining only four
building blocks: buffer, NOT, constant, and branch \cite{49}.
Together with AQFP latch \cite{AQFPlatch} these blocks enable design
of adiabatic circuit of arbitrary complexity. Recently, 10 thousand
gate-scale AQFP circuit was reported \cite{AQFP10K}.

Magnetic coupling of AQFP gates is performed via transformers.
Current flowing through the transformer wire cannot be too small
because it ought to provide appropriate bias flux to subsequent cell
despite of possible technological spread of AQFP parameters. This
limits maximum wire length to about $1$~mm \cite{49}. This length is
further conditioned by trade-off with maximum clock frequency, which
is limited to $5$~GHz in practical circuits
\cite{48,45,47,46,AQFPlatch,49,50}. This clock frequency limitation
relaxes complexity of AC bias lines design. However, with circuit
scale increase, lengthy distribution of clock lines is nonetheless
expected to generate a clock skew between logic cells
\cite{AQFPrec}.

While adiabatic circuits are clearly the most energy efficient ones,
their operation frequency is  relatively low and the latency is
relatively large. However, recently it was shown that due to
intrinsic periodicity of AQFP potential energy, the cell can be
operated at double or even quadruple activation current frequency
with an increase of the current amplitude \cite{MEAQFP}. This opens
oportunity to speed up AQFP circuits up to $10$~GHz or even $20$~GHz
clock.

\subsubsection{nSQUID-based circuits}

Above, we already mentioned that it is possible to swap the roles of
activation current and input/output in the parametric quantron. In
this case, information is represented in magnetic flux of the SQUID,
while its bias current flowing through the main parametric quantron
loop plays the role of excitation.

It was noted that while the SQUIDs of different such cells may be
coupled magnetically, their activation current pulse can be provided
sequentially using a common bias bus. For better control of the
SQUID state in this scheme the value of the main parametric quantron
loop inductance should be minimized. In addition, it was proposed to
provide negative mutual inductance between the two parts of the
SQUID loop inductance \cite{41}. SQUID with negative mutual
inductance is called ``nSQUID''. Its inductance is effectively
decreased for the bias current but increased for the current
circulating in its loop.

\begin{figure}
\includegraphics[bb=0 0 340 194,width=1\columnwidth,keepaspectratio]{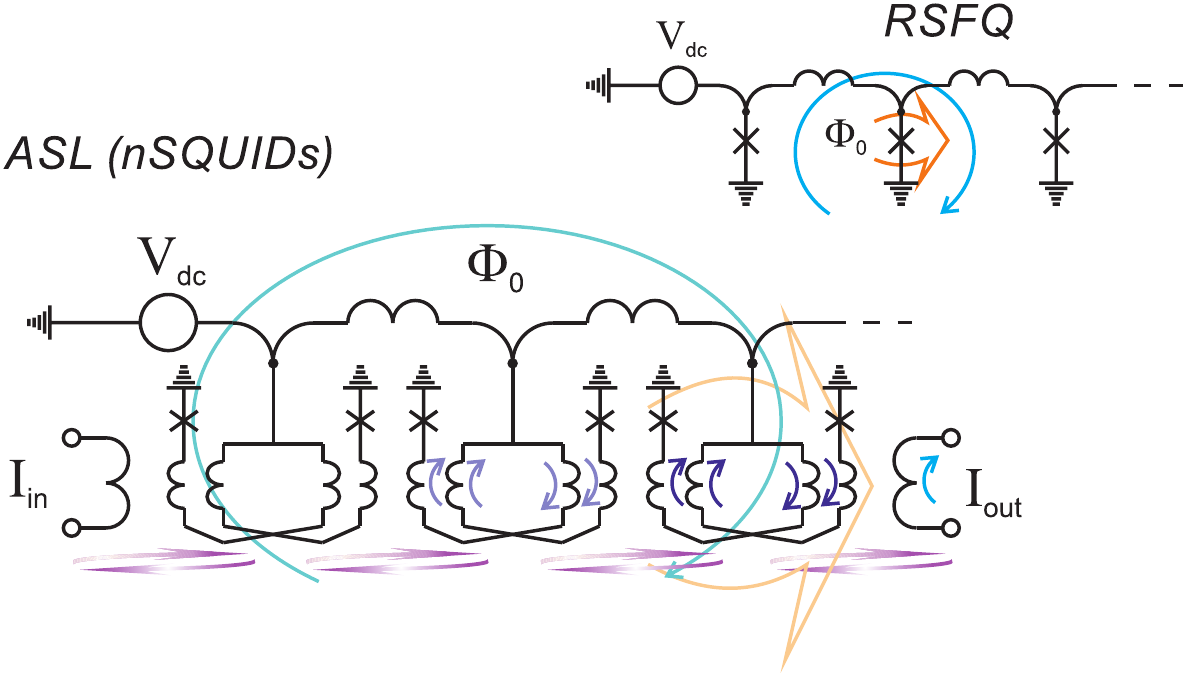}
\caption{nSQUID-based adiabatic data bus and RSFQ data bus. Blue
arrows show circulating currents, orange arrows highlight critically
biased elements, violet arrows present magnetic coupling.}
\label{Fig13}
\end{figure}

Successive application of activation current pulse to nSQUIDs from a
common bias bus can be realized by using an SFQ \cite{41,42,43,44}.
Note that nSQUID-based transmission line is quite similar to
conventional RSFQ JTL with substitution of Josephson junctions by
nSQUIDs, see Figure~\ref{Fig13}. Here data bit is spatially bound to
SFQ. Such application of activation current pulse allowed to switch
from AC to DC power supply. It was shown that it is possible to
switch also from magnetic to galvanic coupling between nSQUIDs
\cite{42}.

nSQUID-based circuits were successfully tested \cite{43} at $5$~GHz
clock frequency. At lower frequency, $50$~MHz, their energy
dissipation per logic operation was estimated \cite{8} to be close
to the thermodynamic limit, $\sim 2 k_B T\ln 2$.

Since nSQUID circuits utilize SFQ clocking, the clock rate (and
hence, the power dissipation) can be adjusted ``in flight'' like in
RSFQ circuits. Note that the energy associated with SFQ creation or
annihilation $E_J$ is much greater than the thermodynamic limit at
$4.2$~K. SFQs are ``recycled'' to avoid this energy dissipation. For
this purpose the circuits are made in closed loop manner as ``timing
belts'' \cite{44}. Thus, total number of SFQs remains unchanged.
However, this imposes certain restriction on the circuit design.

It is interesting to note that it was proposed to use nSQUID
circuits for implementation of ``flying qubits'' transmitting
quantum information \cite{42,FLQubit}. Yet, this idea is not
implemented experimentally.

\subsection{Discussion}

We considered non-adiabatic and adiabatic logics which
implementations are different mainly in the type of power supply, AC
or DC. Each type has its own advantages and disadvantages.

The most attractive feature of AC versus DC bias is that the power
is supplied in series. We should note that this feature can be
utilized also for DC biased circuits by using AC-to-DC converter
\cite{ACDCconv}. At particular frequency of AC power source the
required bias voltage can be obtained by serial connection of these
converters. Power supply of different parts of large scale DC biased
circuit by such voltage sources could eliminate the need for the
circuit partitioning.

In general, SFQ AC biased circuits are good in design of large
regular structures. The largest superconductor digital circuit is AC
biased shit register containing 809 thousand Josephson junctions
\cite{RQL_SR}. It was used as a fabrication process benchmark
circuit like a kind of ``scan chain''.

DC power source is most convenient in terms of providing the power
into the cryogenic system. Indeed, the bandwidth of microwave cables
is often narrow to prevent hit inflow. In order to overcome the
limitation on the maximum frequency of AC biased circuits it was
proposed to use a DC-to-AC converter as on-chip power source
\cite{DCAC}. This converter was successfully tested in experiment
providing oscillation frequency of $4.4$~GHz. The output AC bias
current amplitude can be tuned by varying DC bias current of the
convertor. Utilization of AC-to-DC and DC-to-AC converters allows to
use circuits based on different logics on a single chip, increasing
the variability of design.

Physical localization of information corresponding to quantization
of magnetic flux leads to another issue, especially in digital SFQ
circuits. Due to low gain from Josephson junctions, the circuits are
featured by low fan-out. An SFQ has to propagate through large and
slow SFQ splitter tree to split information into multiple branches.
The same situation is with merging of multiple outputs.

Solution of this problem can be found in utilization of magnetic
control over cells by using current control line. This approach can
be realized with SFQ-to-current loop converter
\cite{CLD,RSFQdecoder}. Similar technique can be used in merging of
multiple outputs \cite{SQFMerg}.

SFQ-to-current conversion can be realized also by
Superconducting-Ferromagnetic Transistor (SFT) \cite{SFTNev} or by
``non-Josephson'' device like n-Tron \cite{nTron1}. The former is
the three (or four) terminal device comprising two stacked Josephson
junctions. One of them, ``injector'', (containing ferromagnetic
layer(s)) serves for injection of spin-polarized electrons in common
superconducting electrode of both junctions, thus suppressing its
superconductivity. This manifests itself as redistribution of
superconducting current flowing through this electrode or as
degradation of ``acceptor'' (typically SIS junction) critical
current depending on configuration of the device \cite{SFTNev1}.
While having good input/output isolation, SFT is capable of
providing voltage, current, and power amplification.

n-Tron is the three terminal device comprising superconducting strip
with a narrow in the middle to which the third terminal tip is
connected. Current pulse from the third terminal switch off
superconductivity of the nanowire, that is similar to SFT operation
to some extent. Unlike Josephson junction, the nanowire in resistive
state possesses several $k\Omega$ resistance which provide high
output impedance and high voltage signal \cite{nTron2,nTron3}. Both
devices can be utilized as an interface
\cite{SFTNev1,nTronCMOS1,nTronCMOSmem} between superconductor
circuit and CMOS electronics or memory depending on requirements to
output signal and energy efficiency.

It is well known that the major computation time and power
consumption is associated with communications between logic and
memory circuits \cite{MAGIC2}. Logic cells possessing feature of
internal memory are now being considered as possible element base
for development of new, more efficient computers \cite{Mec1,Mec2}.
Superconductor logic circuits utilizing their internal memory were
named ``MAGIC'' (Memory And loGIC) circuits \cite{MAGIC2,MAGIC1}.
This concept is based on conventional ERSFQ cells involving their
renaming or rewiring. It promises an increase in clock rate to above
100 GHz threshold, combined with up to ten-fold gain in functional
density. In general, the mentioned quantum localization of
information and high non-linearity of Josephson junctions make
superconductor circuits to be ideally suit for implementation of
unconventional computational paradigms like cellular automata
\cite{CellA1,CellA2}, artificial neural networks
\cite{ANN1,ANN2,ANN3} or quantum computing
\cite{QCR1,QCR2,QCRIF1,QCRIF2,QCRIF3}.

Unfortunately, the major problem of superconductor circuits does not
relate to a particular logic of computation. Low integration density
in all cases limits complexity, and therefore performance of modern
digital superconductor device. Possible solutions here are
miniaturization of existing element base and increase of its
functionality.

The first one can be performed by scaling down the SIS Josephson
junction \cite{250nmJJ}, or search for other high accuracy
technological processes providing nanosized junctions with high
critical current density and normal-state resistance
\cite{250nmJJ,aSi,aSi1}. Another direction of the research is
substitution of conventional loop inductance for kinetic inductance
or inductance of Josephson junction \cite{T}. This also allows to
make the circuits more energy efficient. Indeed, Josephson junction
critical current $I_c$ and loop inductance $L$ are linked for SFQ
circuits. Their product should be $I_c L \approx \Phi_0$ for proper
operation. While the critical current $I_c$ has to be decreased in
order to improve the energy efficiency, $E_J \approx I_c \Phi_0$,
this leads to increase in the inductance making the circuit to be
sparse. Miniaturization of inductance weakens this problem.
Unfortunately, transformer remains an inherent component of the
circuits which can not be miniaturized in this way.

One should note that contrary to CMOS technology where transistor
layer is implemented on a substrate, Josephson junctions can be
fabricated at any layer. This provides opportunity for utilization
of 3D architecture. With planned technological advances, the
Josephson junction density up to $10^8/$cm$^2$ seems achievable.

Finally, Josephson junctions with unconventional current-phase
relation (CPR) can be utilized in a circuit for its miniaturization.
For example, the so-called ``$\pi$''-junction (the junction with
constant $\pi$ shift of its CPR) can be used as a ``phase battery''
providing constant phase shift \cite{pJJ1,pJJ2} instead of
conventional transformer. Control of the junction CPR phase shift
\cite{LUTwFJJ} can provide the change in the logic cell functioning,
e.g., converting AND to OR. This mechanism can be also used for
implementation of memory cell \cite{LUTwFJJ,NGmemcell}.

Historically, the problem of element base miniaturization was first
recognized in development of superconductor random access memory
(RAM). Since that time, the need for dense cryogenic RAM is the
major stimulus for innovative research in this area.

\section{Memory}

Among the many attempts to create a cryogenic memory compatible with
energy-efficient superconducting electronics, we want to single out
the four most productive competing directions: (A) SQUID-based
memory, (B) hybrid Josephson-CMOS memory, (C) JMRAM and (D)
OST-MRAM.

\subsection{SQUID-based memory}

The presence or absence of SFQ(s) in a superconducting loop can be
the physical basis for a digital memory element. Due to high
characteristic frequency of Josephson junction, SQUID-based memory
cells stand out with fast (few picoseconds) \cite{BLZ} write/read
time favorable for RAM which is indispensable for data processor.
Throughout various SQUID-based RAM realizations memory element was
provided with destructive \cite{MDRO1,MDRO2,MDRO3} or
non-destructive \cite{MNDRO1,MNDRO2,MNDRO3} readout. Memory cell
contained accordingly from two \cite{MDRO1} to ten \cite{BLZ}
Josephson junctions. With Josephson junction micron-scale dimensions
in the late 1990-s this resulted in memory cell area of an order of
few hundreds of microns squared. While power dissipation per
write/read operations was at $\mu$W level, memory chip capacity
\cite{MNDRO3} was only up to $4$~kb. In the particular $4$~kb RAM
memory \cite{MNDRO3} the memory drivers and sensing circuits
required AC power which limited its clock frequency to $620$~MHz.
Later, all-DC-powered high-speed SFQ RAM based on pipeline structure
for memory cell arrays was proposed \cite{MNDRO4}. Estimation showed
that this approach allows up to $1$~Mb memory on $2\times2$~cm$^2$
chip operated at $10$~GHz clock frequency and featuring $12$~mW
power dissipation. Still, it was never realized in experiment.

\subsection{Hybrid Josephson-CMOS memory}

Low integration density of SQUID-based memory cells seemed to be
significant obstacle to the development of low-temperature RAM with
reasonable capacity. This approach was succeeded by hybrid
Josephson-CMOS RAM where Josephson interface circuits were amended
by CMOS memory chip \cite{HMEM1,HMEM2,HMEM3,HMEM4,HMEM5}. This
combination allowed to develop $64$~kb, $4$~K temperature RAM with
$400$~ps read time, and $21/12$~mW power dissipation for write/read
operations, accordingly \cite{HMEM5}. CMOS memory cell was composed
of 8 transistors. While being fabricated in a $65$~nm CMOS process,
the cell size was about three orders of magnitude less than the one
of its SQUID-based counterparts. Main challenge in design of this
memory system was amplification of sub-mV superconductor logic
signal up to the $\sim 1$~V level required by CMOS circuits. This
task was accomplished in two stages. First, the signal was amplified
to $60$~mV using a Suzuki stack, which can be thought as a SQUID
with each Josephson junction substituted by a series array of
junctions for high total resistance \cite{SuzukiS}. At the second
step, the reached $60$~mV signal drives high sensitive CMOS
comparator to produce the volt output level.

Suzuki stack \cite{HMEM4} and CMOS comparator \cite{CMOScomp} were
optimized for best compromise of power and time performance. Their
simulated power $\times$ delay product for read operation were
$2.3$~mW $\times$ $47$~ps ($0.11$~pJ) and $6.4$~mW $\times$ $167$~ps
($1.1$~pJ). This made up $73$\% and $53$\% of total memory system
read power and time delay, correspondingly. These results exhibit
severe restriction of overall system performance by the interface
circuits. Recently, it was shown that the power consumption can be
significantly decreased with utilization of energy efficient ERSFQ
decoders and n-Trons as high voltage drivers \cite{nTronCMOSmem}.
This could provide the energy efficiency improvement up to $3$ times
for $64$~kb, and up to $12$ times for $16$~Mb memory. In the latter
case, the access time in a read operation is estimated to be
$0.78$~ns.

While the hybrid memory approach showed better memory capacity, its
power consumption and time requirements are still prohibitive. It
was summarized that for implementation of practical low-temperature
RAM one should meet the following criteria \cite{MRep}: (i) scale:
memory element dimension $< 100$~nm ($< 200$~nm pitch); (ii) write
operation: $10^{-18}$~J energy with $\sim 50 - 100$~ps time delay
per cell; (iii) read operation: $10^{-19}$~J energy with $\sim 5$~ps
time delay per cell. An idea to meet the requirements nowadays is to
bring spintronics (including superconductor spintronics) in RAM
design.

\subsection{JMRAM}

It is possible to reduce drastically the size of superconducting
memory cell by using controllable Josephson junction with magnetic
interlayers instead of SQUID
\cite{JMRAM1,JMRAM2,JMRAM3,JMRAM4,JMRAM5,JMRAM6}. Topology of such
magnetic Josephson junction (MJJ) is usually of two types: (i)
sandwich topology which is well suited for CMOS-compatible
fabrication technology, and (ii) the one with some heterogeneity of
the junction weak-link area in-plane of the layers. Below we present
MJJ valves according to this classification.

\subsubsection{MJJ valve of sandwich topology}

\begin{figure}
\includegraphics[bb=0 0 145 95,width=0.65\columnwidth,keepaspectratio]{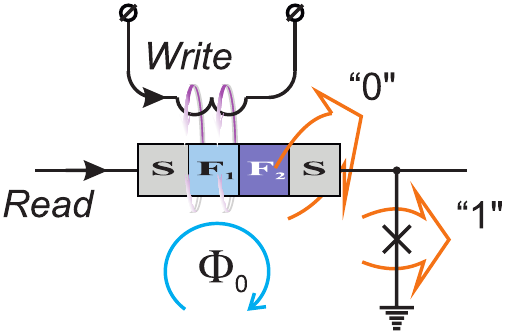}
\caption{Principal scheme of implementation of write and read
operations in MJJ valve based circuit.} \label{Fig14}
\end{figure}

Search for an optimal way of compact MJJ valve implementation
remains under way now. The most obvious solution is to use two
ferromagnetic layers with different magnetic rigidity in the area of
the junction weak link \cite{JMRAM7,JMRAM8,JMRAM9}. Critical current
of such junction is determined by effects resulting from coexistence
and competition of two orderings for electron spins:
``superconducting'' (S) (with usually antiparallel spins of
electrons in the so-called ``Cooper pairs'') and ``ferromagnetic''
(F) (with parallel ordering of electron spins). Magnetization
reversal of ``weak'' F-layer leads to switching between collinear
and anti-collinear orientations of the F-layers magnetic moments in
the bilayer. This, in turn, provides alteration in the total
effective exchange energy, $E_{ex}$,  and hence, in MJJ critical
current effective suppression. While magnetization reversal can be
executed by application of an external magnetic field
\cite{JMRAM10}, the critical current can be read-out, e.g., with MJJ
inclusion into decision making pair \cite{JMRAMRyaz}, see
Figure~\ref{Fig14}. It is possible to trace some analogy between
this effect and the phenomenon of giant magnetoresistance
\cite{JMRAM12} which is actively used in conventional magnetic
memory cells.

A common drawback of most MJJs is small value of their
characteristic frequency ($\omega_c \sim I_c R_n$) in comparison
with SIS junction. Indeed, here one has to perform the magnetization
reversal of weak F-layer with relatively small exchange energy in
order to manipulate the total critical current against the
background of its considerable suppression by the strong
ferromagnet. Low $\omega_c$ outflows in slow read operation and
complicates MJJ integration in SFQ logic circuits.

There are several approaches to solve this problem. One of them is
the using of noncollinearly magnetized ferromagnetic layers
\cite{JMRAM13,JMRAM14,JMRAM15,conclmem1}. In this case, the triplet
superconducting correlations of electrons are formed in the junction
weak-link area. A part of them are featured by collinear orientation
of electron spins in Cooper pairs. They are unaffected by exchange
field of the ferromagnets, thus increasing MJJ critical current
while maintaining its normal state resistance. The ``triplet''
current can be controlled by external magnetic field through
magnetization reversal \cite{Birgereq}. Still, this approach implies
implementation of a number of additional layers (and interfaces) in
the structure that reduces its critical current.

One should also note that $E_{ex}$ alteration could result in $\pi$
shift of MJJ CPR. In this case the valve can be utilized as
controllable phase battery \cite{LUTwFJJ}. Inclusion of such MJJ
into SQUID loop allows fast read-out of its state \cite{NGmemcell}.
However, here miniaturization reduces only to replacement of the
SQUID inductance by the MJJ.

Another approach is based on localization of magnetic field source
outside Josephson junction weak-link area but in the nearest
proximity \cite{FFoutJJ1,FFoutJJ2,FFoutJJ3,FFoutJJ4}. For example,
F-bilayer can be placed on top of SIS junction. In this case, stray
magnetic field penetrating into the junction area controls its
critical current. If the junction S-layer neighboring the F-bilayer
is thin enough, the coupling of the vector potential of the  stray
magnetic field to superconducting order parameter phase could also
noticeably affect Josephson phase difference across SIS junction.
SIS junction is utilized here just for read-out the ferromagnet
state, and therefore, its characteristic frequency remains high.
Still, since the strength of magnetic field is proportional to the
ferromagnet volume, a possibility of miniaturization of such memory
element is doubtful.

Critical current modulation can be obtained even in the structure
with a single magnetic layer by changing the value of its residual
magnetization \cite{JMRAM16}. It is also possible to improve the
characteristic frequency by the inclusion of dielectric (I) and thin
superconducting (s) layers in MJJ weak-link area to increase $R_n$
and $I_c$, correspondingly
\cite{JMRAM19,JMRAM20,JMRAM21,JMRAM22,JMRAM23,JMRAM24,JMRAM25,JMRAM26}.
Such SIsFS valves possess characteristics close to SIS junction
\cite{JMRAM21}. However, compatibility with superconductors requires
utilization of ferromagnets with relatively low coercive field,
which are typically characterized by non-square shape of the
hysteresis loop. This in turn outflows into uncertainty of MJJ
critical current at zero applied magnetic field after multiple
magnetization reversals. In addition, miniaturization here faces the
same difficulties as in the previous approach. For these reasons, it
seems especially fruitful to replace the I and F layers with two
magnetic insulator IF-layers to construct a Josephson S(IF)s(IF)S
valve \cite{JMRAM27,JMRAM28,JMRAM29}. Its operation relies on
variable suppression of superconductivity of the middle s-layer.
Yet, this promising structure is complicated for fabrication.

\subsubsection{MJJ valve with in-plane heterogeneity of the weak-link
area}

\begin{figure}
\includegraphics[bb=0 0 240 162,width=0.6\columnwidth,keepaspectratio]{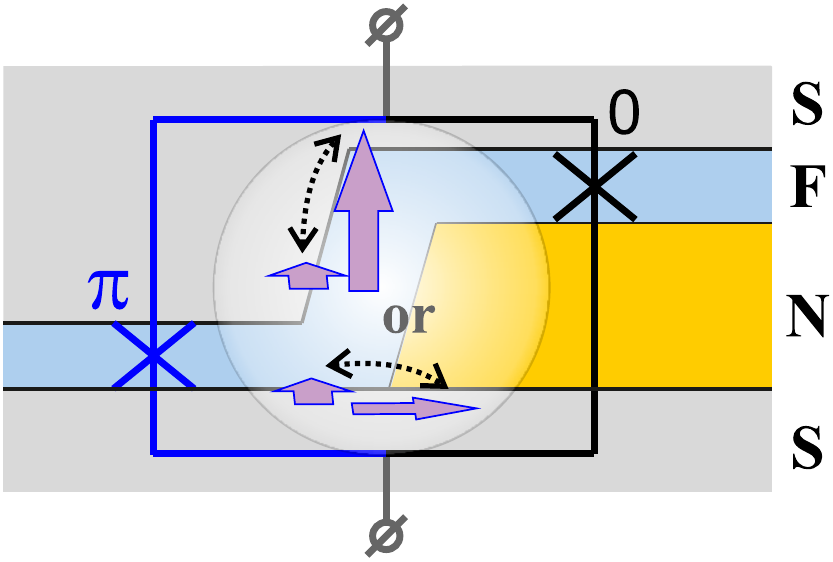} \caption{Cross
section of SF-NFS MJJ with CPRs of its parts shifted in phase by
$\pi$. Arrows show F-layer magnetization directions corresponding to
the valve on/off states.} \label{Fig15}
\end{figure}

The second type of valves implies heterogeneity of the weak-link
region in the junction plane providing separation of the structure
into two parts. CPR of these parts can be different, e.g., the
conventional CPR and the ones shifted in phase by $\pi$
\cite{JMRAM24,JMRAM18}. Such MJJ may be thought as nanoSQUID with
conventional and ``$\pi$'' lumped junctions. Its implementation may
comprise a ferromagnetic interlayer and a sandwich containing the
same F-layer and a normal metal (N) layer, see Figure~\ref{Fig15}.
If F-layer magnetization is aligned perpendicular to the nanoSQUID
plane, it compensates the Josephson phase gradient across the MJJ
making its critical current to be high. While the magnetization is
being rotated at $90^\circ$, this effect is turned off and $I_c$
becomes low. Here for proper operation of this SF-NFS-based MJJ the
flux of residual magnetization must be comparable with the flux
quantum $\Phi_0$.

The second common problem of MJJ-based memory is a long time of
write operation. Bit write is commonly performed by magnetization
reversal of at least one of F-layers. For this reason, recording
time is of an order of inverse frequency of ferromagnetic resonance.
It is usually more than two orders of magnitude larger than the
characteristic time of SIS junction switching. Thus, elimination of
magnetization reversal from the valves operation is desired. It is
worth noting that nano-sized trap for single Abrikosov vortex in the
vicinity of Josephson junction \cite{JMRAM33,JMRAM34} allows to
realize fast enough write operation. However, energy dissipation
associated with annihilation of such vortex ($\sim 10^{-18}$~J) may
contradict with the paradigm of energy-efficiency.

This challenge can be met with MJJ having bistable Josephson
potential energy. Josephson phases of its ground states could be
equal to $\pm \varphi$ ($0 < \varphi < \pi$). One can realize both
write and read operations with such $\varphi$-junction on picosecond
timescale
\cite{JMRAM35,JMRAM36,JMRAM37,JMRAM38,JMRAM381,JMRAM39,JMRAM40}. The
disadvantage of this approach is the difficulties with
$\varphi$-state implementation. In practice, it is possible only in
the structure with heterogeneous weak-link region of a rather large
size.

One more operation principle of MJJ valves relies on control of
superconducting phase domains formation \cite{JMRAM26}. The effect
can be realized in SIsFS MJJ with sFS part substituted, e.g., for
heterogeneous SF-NFS  combination. The middle s-layer is broken on
domains with different superconducting phases if Josephson phases of
the structure parts are different, and vice versa. This process can
be controlled by current injection through sFS or sFNS parts. The
domain formation significantly changes the MJJ critical current.
This MJJ provides fast read and write operations with no need for
application of external magnetic field. Still, fabrication of
compact Josephson junctions having the inhomogeneous weak-link
region with reproducible characteristics is a difficult
technological task.

\subsection{OST-MRAM}

The next considered type of cryogenic memory is the hybrid approach
combining superconducting control circuits with spintronics memory
devices. Here due to spin-based interactions between atoms in the
crystal lattice and electrons, orientations of ferromagnets
magnetization can determine the amount of current flow. And vice
versa, spin-polarized current can affect orientations of the
magnetizations. The last effect is the so-called ``spin transfer
torque'' (STT). It was suggested as a control mechanism for magnetic
memory \cite{COST1,COST2,COST3}. However, high speed and low energy
of write operation can not be provided with conventional spin-valve
topology with collinear orientations of ferromagnets magnetizations
\cite{COST4}.

\begin{figure}
\includegraphics[bb=0 0 89 58,width=0.55\columnwidth,keepaspectratio]{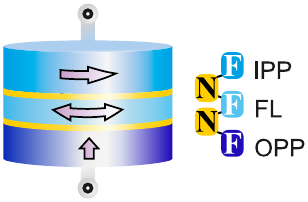}
\caption{Sketch of OST device. Arrows show magnetization directions
in the device layers.} \label{Fig16}
\end{figure}

Orthogonal spin transfer (OST) device allows to overcome the
difficulties. This structure consists of an out-of-plane
ferromagnetic polarizer (OPP), a free F-layer (FL), and a fixed F
in-plane polarizer/analyzer (IPP), see Figure~\ref{Fig16}. ``Write''
current pulse passing through OPP provides STT effect in FL which
acts to lift its magnetization out of plane. Magnetization is then
rotated about the out-of-plane axis, according to
Landau-Lifshitz-Gilbert equation. Current pulse applied to IPP
read-out collinear or anti-collinear magnetizations of in-plane
magnetized F-layers.

It is possible to obtain the necessary $180^{\circ}$ magnetization
reversal with correct selection of the write pulse amplitude and
duration. Quasi-static and dynamic switching characteristics of OST
devices have been analyzed at cryogenic temperatures: switching
between parallel and anti-parallel spin-valve states has been
demonstrated for $\sim$~mA current pulses of sub-ns duration
\cite{COST4,COST5,COST6}.

Clear advantages of the considered approach is elimination of
control lines for magnetic field application, and implementation of
fast magnetization reversal at sub-ns timescale. At the same time,
the problems like relatively low percent of magnetoresistance, and
the ones associated with possible magnetization over-rotation still
prevent its practical application. The latter one can be overcome to
some extend by involving both IPP and OPP polarizers into FL
switching process \cite{COST4}.

One could note that application of STT effect in some of MJJ valves
is of considerable interest. STT in voltage biased superconducting
magnetic nanopillars (SFNFS and SFSFS junctions) has been studied
for both equilibrium and nonequilibrium cases
\cite{COST9,COST10,COST11,COST12,COST13}. However, rich dynamics
resulting from interplay of multiple Andreev reflection, spin
mixing, spin filtering, spectral dynamics of the interface states,
and the Josephson phase dynamics requires further research for
evaluation of STT application appropriateness in superconducting
memory structures.

\subsection{Discussion}

Lack of suitable cryogenic RAM is ``... the main obstacle to the
realization of high performance computer systems and signal
processors based on superconducting electronics.'' \cite{MEMDIS1}
While JMRAM and OST-MRAM look as the most advanced approaches, they
still require further improvement in a number of critically
important areas.

Progress in considered variety of device types with no clear winner
is impossible without researches on new magnetic materials like
PdFe, NiFe(Nb,Cu,Mo), Co/Ru/Co, [Co/Ni]$_n$ etc., and novel
magnetization reversal mechanisms \cite{MEMDIS2,MEMDIS3,MEMDIS4}.
They can lead to development of new operation principles combining
superconductivity and spintronics.

Inverse proximity effect at SF boundaries dictates utilization of
pretty thin (at nm scale) magnetic layers. However, characteristics
of memory devices typically depends exponentially on the F-layers
thicknesses and significantly affected by interfacial roughness.
This challenge can be met with further development of high-accuracy
thin-film technological processes in modern fabrication technology.

Substantial part of circuit area, time delay and dissipated power in
memory matrix is more likely to be associated with address lines
rather than with memory cells. This makes optimization of
intra-matrix interconnections and memory cell architecture of
significant importance.

While we considered here only the most developed solutions for
superconducting valves and memory elements, there are many other
approaches to create nanosized controllable superconducting devices
for applications in memory and logic. We can point out on our
discretion: the nanoscale superconducting memory based on the
kinetic inductance \cite{MEMDIS5}, and the superconducting quantum
interference proximity transistor \cite{MEMDIS6}. Such concepts
could bring novel idea into nanoscale design of superconducting
circuits.

\section*{Conclusion}

In conclusion, we discussed different superconductor logics
providing fast ($\sim 5 - 50$~GHz) and energy efficient ($10^{-19} -
10^{-20}$~J per bit) operation of circuits in non-adiabatic and
adiabatic regimes. The last one allows implementation of the most
energy efficient physically and logically reversible computations
with no limit for minimum energy dissipation per logic operation.
Possibilities to combine the schemes based on different logics as
well as utilization of different (e.g. superconductor and
semiconductor) technologies in a single device design are presented.

Considered physical principles underlying superconducting circuits
operation provide possibility for development of devices based on
unconventional computational paradigms. This could be the basis for
a cryogenic cross-platform supercomputer, where each task can be
executed in the most effective way. In our opinion, the development
of superconducting circuits performing non-classical computations
like cellular automata, artificial neural networks, adiabatic,
reversible, and quantum computing is indispensable to get all the
benefits of the superconductor technology.

Low integration density, and hence low functional complexity of the
devices, is identified as the major problem of the considered
technology. This issue can be addressed with further miniaturization
of basic elements and modernization of cell libraries, including
introduction of novel devices like the ones based on nanowires or
magnetic Josephson junctions.

The problem of low integration density is especially acute in RAM
design. We considered here four different approaches to cryogenic
RAM development with no clear winner. Progress in this area now
implies elaboration of new operation principles based on synergy of
different physical phenomena like superconductivity and magnetism,
and appearance of novel effects, as for example, triplet spin valve
memory effect \cite{conclmem1} or superconducting control of the
magnetic state \cite{JMRAM27}. Proposed concepts of new controllable
devices could eventually change the face of superconductor
technology making it universal platform of future high-performance
computing.

\begin{acknowledgements}
The authors are grateful to Alexander Kirichenko and Timur Filippov
for fruitful discussions. This work was supported by grant No.
17-12-01079 of the Russian Science Foundation, Anatoli Sidorenko
would like to thank the support of the project of the Moldova
Republic National Program "Nonuniform superconductivity as the base
for superconducting spintronics" ("SUPERSPIN", 2015-2018), partial
support of the Program of Competitive Growth at Kazan Federal
University is also acknowledged.
\end{acknowledgements}

\end{document}